\documentclass[twocolumn]{aa}  

\usepackage{graphicx}
\usepackage{natbib}
\usepackage{color}           
\definecolor{mgc}{RGB}{0,0,192}

\begin{document}

   \title{Plasma motions and non-thermal line broadening in flaring twisted coronal loops}

   \author{M. Gordovskyy \inst{1}\fnmsep\thanks{\email{mykola.gordovskyy@manchester.ac.uk}}, 
	   E.P. Kontar\inst{2}
          \and
          P.K. Browning\inst{1}}

   \institute{Jodrell Bank Centre for Astrophysics, University of Manchester, Manchester M13 9PL, UK\\
         \and
             Astronomy \& Astrophysics Group, University of Glasgow, Glasgow G15 8QQ, UK\\
             }

   \date{Received ; accepted }

\authorrunning{Gordovskyy et al.}
\titlerunning{Plasma motions in twisted loops}

 
  \abstract	
   {Observation of coronal EUV spectral lines sensitive to different temperatures offers an opportunity to evaluate the thermal structure and flows in flaring atmospheres. This, in turn, can be used to estimate the partitioning between the thermal and kinetic energies released in flares. }
   {Our aim is to forward-model large-scale (50-10000~km) velocity distributions in order to interpret non-thermal broadening of different spectral EUV lines observed in flares. The developed models allow us to understand the origin of the observed spectral line shifts and broadening, and link these features to particular physical phenomena in flaring atmospheres.}
   {We use ideal MHD to derive unstable twisted magnetic fluxtube configurations in a gravitationally-stratified atmosphere. The evolution of these twisted fluxtubes is followed 
	using resistive MHD, with anomalous resistivity depending on the local density and temperature. The model also takes into account the thermal conduction and radiative losses in the continuum.
	The model allows us to evaluate average velocities and velocity dispersions, which would be interpreted as ``non-thermal'' velocities in observations, at different temperatures for different parts of the models.}
   {Our models show qualitative and quantitative agreement with observations. Thus, the line-of-sight (LOS) velocity dispersions demonstrate substantial correlation with the temperature, increasing from about 20-30~km/s around 1~MK 
to about 200-400~km/s near 10-20~MK. The average LOS velocities also correlate with velocity dispersions, although they demonstrate a very strong scattering, compared to observations. We also note that near foot-points the velocity dispersions across the magnetic field are systematically lower than that along the field. We conclude, that the correlation between the flow velocities, velocity dispersions and temperatures are likely to indicate that the same heating mechanism is responsible for heating the plasma, its turbulisation and expansion/evaporation.}
   {}


   \maketitle
%

\section{Introduction}\label{intro}

\begin{figure*}[ht!]    
\centerline{\includegraphics[width=0.65\textwidth,clip=]{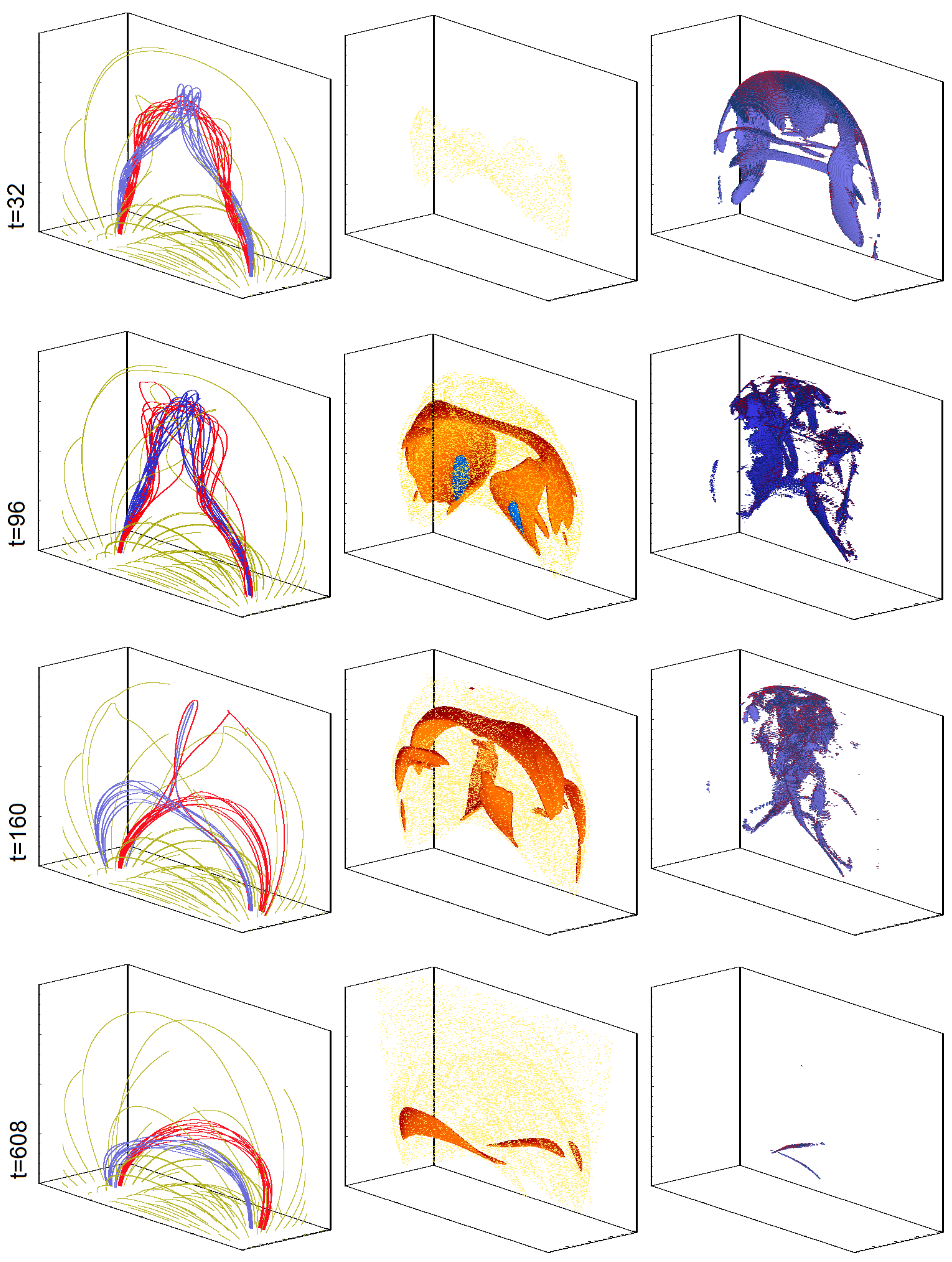}}\label{f-mhdloop}
\caption{Evolution of {\it Model L1}. {\it Left column:} Magnetic field lines; red and blue field lines are tied to opposite footpoints of the twisted loop, yellow lines denote ambient field. {\it Middle column:} Temperature distribution; yellow patchy 
surface corresponds to 2~MK, orange and blue surfaces correspond to 3 and 4~MK, respectively. {\it Right column:} Surfaces of $j=j_{crit}$, i.e. showing the locations with fast magnetic energy release. 
Times (after onset of the instability) are shown for each row on the left in units of characteristic Alfven times $t_A$, which are $0.44$~s for this model.}
\end{figure*}

Plasma motions in the flaring solar corona are closely related to heating processes and can be used to study the spatial distribution of energy release and dynamics of energy transfer. 
New instruments onboard recent solar space missions, such as Hinode/EIS and SDO/AIA, offer an 
opportunity to study characteristics of very hot ($\sim$1-10~MK and hotter) plasma in flaring atmosphere with high spatial resolution \citep{cule07,leme12}. 
Observations show that solar flares have a complex plasma flow structure at different spatial scales \cite[e.g.][]{wate10,dele11}.
Furthermore, flare kernels often reveal coronal lines with complex Doppler structure, containing both red-shifted and blue-shifted components \cite[e.g.][]{youe13}, indicating that plasma motions are inhomogeneous in spatially unresolved volumes.

Analysis of the Doppler shift structure in coronal lines shows that colder plasma (T~$\approx 0.1-1$~MK) normally moves downwards with velocities of about 20-50~km~s$^{-1}$, while the hotter plasma (T~$\approx 1-10$~MK) shows upflow; this upflowing plasma is very inhomogeneous, showing velocities from only few km~s$^{-1}$ up to $400-500$~km~s$^{-1}$ \cite[see][]{dose07,delz08,dose08,wate10,dose13,youe13}. It is also noted that the downflow plasma can be substantially denser (n~$\sim 3\times 10^{16}$~m$^{-3}$) than the upflow plasma (n~$\sim 5\times 10^{14} - 10^{15}$~m$^{-3}$), which is, probably, why the downflow is easier to observe than the upflow \cite[e.g.][]{delz08,dose08}. 

Most coronal lines in flares also demonstrate non-thermal broadening due to additional velocity dispersion \cite[e.g.][]{dose07}. The non-thermal velocity appears to correlate with the bulk plasma velocity: for instance, in two flares analysed by \citet{dose08} using {\it Hinode} data, non-thermal velocity dispersion is higher at locations with higher upflow speed: the velocity dispersion is about 25-30~km~s$^{-1}$ when the upflow speed is nearly zero, and is about  50-60~km~s$^{-1}$ when the upflow velocity is 20~km~s$^{-1}$. The non-thermal velocities, which appear early in solar flares \cite[see e.g.][]{hare13}, have been found to correlate also with plasma temperature. The non-thermal velocities have lowest values near 1~MK: from 0 to 20-40~km~s$^{-1}$. At lower temperatures ($0.1-1$-MK) these velocities are higher, about 40-60~km~s$^{-1}$. At the same time, at high temperatures (10-20~MK) they can be as much as 100-130~km~s$^{-1}$. The non-thermal velocities at high temperature vary during the flare, they peak very close to the hard X-ray maximum at about 100~km~s$^{-1}$ (some observers note much higher values, up to 380~km~s$^{-1}$, see e.g. \citet{cire13}), and then during the decay phase of the flare reduce to about 50~km~s$^{-1}$ \citep{suse13}. 

The bulk plasma flows are normally interpreted as a vertical advection of plasmas in stratified atmosphere due to heating in the chromosphere, transition region or lower corona. As far as the non-thermal line broadening is concerned, there are several viable interpretations. Firstly, this can be due to the turbulence, which can occur in a very hot plasma in post-reconnection loops or directly at the primary energy release location. The latter is compatible with the idea that the reconnection should occur in a very turbulent medium, and this turbulence would facilitate the magnetic diffusion \cite[see e.g.][]{lavi99,brla13}. It is also expected that the plasma in flaring coronal loops will be strongly turbulent due to relatively fast heating (by energetic particles, waves, shocks, conduction etc). Secondly, coronal lines can also be broadened due to unresolved inhomogeneity of regular flows (either in the plane perpendicular to LOS, or along LOS). This is possible, particularly, in the lower corona and the chromosphere, which can have small-scale structure (as small as $\sim$100~km, see e.g. \citet{anro12,golo14}). The velocity field can also be highly non-uniform in a relatively small volume occupied by the exhaust from reconnecting current layer \citep[e.g.][]{vrse09,koe10}.

The issue of plasma turbulence in solar flares is intimately connected to the problem of energy partitioning. There are several estimations of the energy partitioning in flares, but they appear to be very different, changing from flare to flare \cite[see e.g.][]{emse04,flee15}. The problem is that all observational estimations appear to depend strongly on the instrument properties \cite[see e.g.][]{benz08}. Furthermore, there is a substantial uncertainty about the total energy of non-thermal particles: it depends on the low-energy part of the spectra, which, in turn, depends on the so-called lower energy cut-off, which is still under discussion \cite[e.g.][]{kone08}. Diagnostics of thermal plasma can also be difficult, particularly of low-temperature plasma, which is less visible in EUV continuum and contributes less to coronal spectral lines. Indeed, an adequate evaluation of the turbulent velocities is necessary in order to determine what part of the energy is released as either kinetic, thermal, magnetic or potential energy. The first two types depend on the plasma flows, turbulence, temperature and density distributions in the plasma, and energetic (i.e. non-Maxwellian) particle spectra. In this study, we derive the parameters of thermal plasma from our MHD models. 

Most studies interpret these observations based on the so-called standard model of solar flares. At the same time, flare models based on magnetic reconnection in kink-unstable twisted coronal loops \citep{broe08,hooe09,gobr12,bare13,gore13,gore14}, which are more relevant to confined flares within single loops, offer a good opportunity to investigate plasma motion in flaring loops at different spatial scales, from 
$\sim$100~km to $\sim$10~Mm. 
One advantage of this scenario is that it offers both configurations with rather localised heating (at the loop-top and/or near foot-points), as well as configurations with heating sources quite uniformly
distributed along flaring loops. Characteristics of plasma motion are defined predominantly by the distribution of plasma heating sources (due to Ohmic dissipation, high-energy particle thermalisation etc) and their temporal evolution, and by thermal conduction and radiative energy losses. 
Hence, even though the considered models are not a universal flare scenario, and don't account for energy transfer by energetic particles, the results obtained with 
these models can be extrapolated to a wide range of flare configurations and scenarios.

In the present paper we study turbulent plasma motions, resulting in non-thermal broadening of spectral lines. Our models make it possible to investigate the velocity field, with spatial scale of $\sim$100~km (which is the grid-step in z-direction), which is not resolved in majority of observations. We derive characteristics, which are direct proxies for measured parameters, and discuss them in context of observational data.

\section{Main features of the considered flare model}\label{model}

\begin{figure}[h!]    
\centerline{\includegraphics[width=0.4\textwidth,clip=]{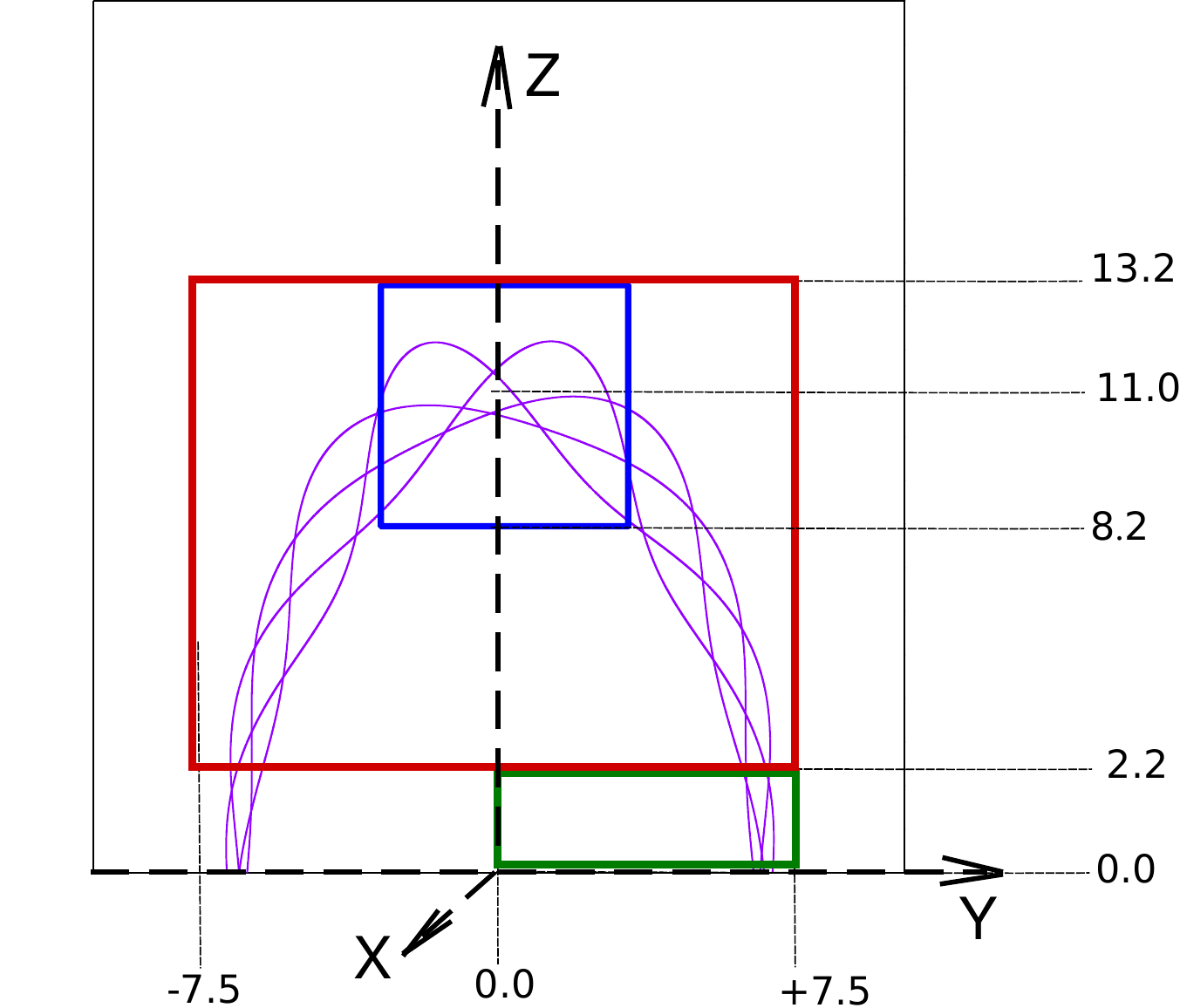}}\label{f-sketch}
\caption{Locations of sampling boxes in the domain: green box denotes the ``footpoint region'', red box denotes the ``coronal region'', and the blue box denotes the ``loop-top region''.
For {\it models S1} and {\it S2} the units are 1~Mm, while for larger {\it models L1} and {\it L2} the units are 4~Mm.}
\end{figure}

We consider four different models of kink-unstable coronal loops with parameters typical for a flaring atmosphere: 
\begin{itemize}

\item {\it Model S1} has a  loop lengths of about 20~Mm with the apex at about 10Mm. Its footpoint magnetic field is 95~G. Field geometry of this loop is very similar to those described by \citep{gore14,pine15}. The model box has dimensions $x=[-10;+10]$~Mm,  $y=[-10;+10]$~Mm, and $z=[0;+20]$~Mm. The axes are shown in Figure~\ref{f-sketch}.

\item {\it Model S2} is considered in the same numerical box and has the same parameters as {\it model S1}, apart from stronger footpoint magnetic field -- it is 200~G.

\item {\it Model L1} has similar configuration to smaller loops ({\it S1} and {\it S2}) but is proportionally larger and has stronger field. Its length is about 80~Mm and footpoint strength is 700~G. The numerical box  is $x=[-40;+40]$~Mm,  $y=[-40;+40]$~Mm, and $x=[0;+80]$~Mm.

\item {\it Model L2} is considered in the same numerical box and has the same parameters as {\it model L1}, apart from stronger footpoint magnetic field -- it is 1500~G.
\end{itemize}
It is assumed that in all the four models the lower boundary corresponds to the same level in the solar atmosphere and, hence, the lower boundary has the same density in all models, $10^{20}$~m$^{-3}$ (approximately corresponding to the upper chromosphere). 

The loops are all embedded into gravitationally stratified atmosphere. The initial density in the atmosphere is 
\begin{equation}
\rho_{t=0}(z) = \rho_1 \exp\left( -\frac {z+z_{sh}}{z_{sc\,1}}\right) + \rho_2 \exp\left( -\frac {z+z_{sh}}{z_{sc\,2}}\right), \nonumber
\end{equation}
where the density constants are $\rho_1=3.34\times 10^{-5}$~kg~m$^{-3}$ and $\rho_2=2\times 10^{-12}$~kg~m$^{-3}$, and the density length scales are $z_{sc\,1}=0.25$~Mm and $z_{sc\,2}=50$~Mm. The shift $z_{sh}$=1.5~Mm, putting the lower boundary of the computational domain approximately at the height of the lower chromosphere. The initial density at the upper boundary is $10^{15}$~m$^{-3}$ in the {\it models S1} and {\it S2}, and $3\times 10^{14}$~m$^{-3}$ in {\it L1} and {\it L2}. The gravity acceleration is assumed to be constant $g=275$~m~s$^{-2}$, so that the initial pressure can be easily obtained from density, as $dp_{t=0}(z)/dz =-\rho_{t=0}(z) g$. As the density constants $\rho_1$ and $\rho_2$, and the length scales $z_{sc\,1}$ and $z_{sc\,2}$ are very different, there are two regions with nearly constant temperatures: around 10$^4$~K below 2~Mm, and about 0.9~MK about 2~Mm \cite[see][]{gore14}. The magnetic field is strongly convergent, i.e. it is about 10 time weaker near the loop apex, compared to the footpoints. This means that the fluxtube cross-section radius near the apex is around 3.2 times larger than that near foot-points.
The modelling has been performed using the LARE3D code by \citet{arbe01}. The simulations are based on the resistive single-fluid MHD, also incorporating Braginskii thermal conduction \citep{brag65} and radiative losses following \citet{klie08}.

Pre-kink configurations have been derived as described in \citet{gore13,bare15,pine15}. Although the initial atmosphere is in the hydrodynamical equilibrium, it is not stationary because of the thermal conduction. The heat flux from the hot corona to colder chromosphere results in noticeable changes of the temperature around the transition region. Moderate heating of the transition region and chromosphere and moderate cooling of the lower corona lead to smoother temperature profile near the transition region \cite[see][]{pine15,bare15}. The atmosphere settles in about $\sim$1000~t$_A$ and only after this the initially potential loop is twisted by footpoint rotation. The twisting is very slow, the maximum linear velocity in small models is about 10$^{-3}$~v$_0$ (or $\sim 3$~km~s$^1$), which is always lower than local Alfven and sound velocities. Hence, during the twisting phase loops undergo sequences of nearly force-free states until kink instability onset. As the result, the loop undergoes a sequence of nearly-force-free states during the twisting. Also, the twisting velocity is well below the observed LOS velocities and velocity dispersions ($\geq 10$~km~s$^1$).

The kink instability occurs when the total twist reaches $6\pi-8\pi$ \cite[see also][for details]{bare15}.
The magnetic reconnection in twisted coronal loops atmosphere after kink 
instability has been described by in many earlier studies \cite[see][]{hopr79,bahe96,brva03,hooe09,gobr12,gore14,bare15}, while the evolution of thermal and non-thermal radiation 
from such systems is considered by \citet{bote12,gore13,gore14,pine15}. Essentially, a loop experiences two different types of reconnection simultaneously: reconnection between field lines of twisted fluxtube, resulting in twist reduction, and reconnection of twisted field lines with the ambient field, resulting in radial expansion of twisted loops (Figure~\ref{f-mhdloop}). The plasma temperature and, hence, thermal emission intensities peak towards the end of fast reconnection, when the twist is considerably lower. As the result, the visible twist in EUV/SXR is much lower than the critical twist required for the kink instability \citep{pine15}. 

\begin{figure*}[ht!]    
\centerline{\includegraphics[width=0.8\textwidth,clip=]{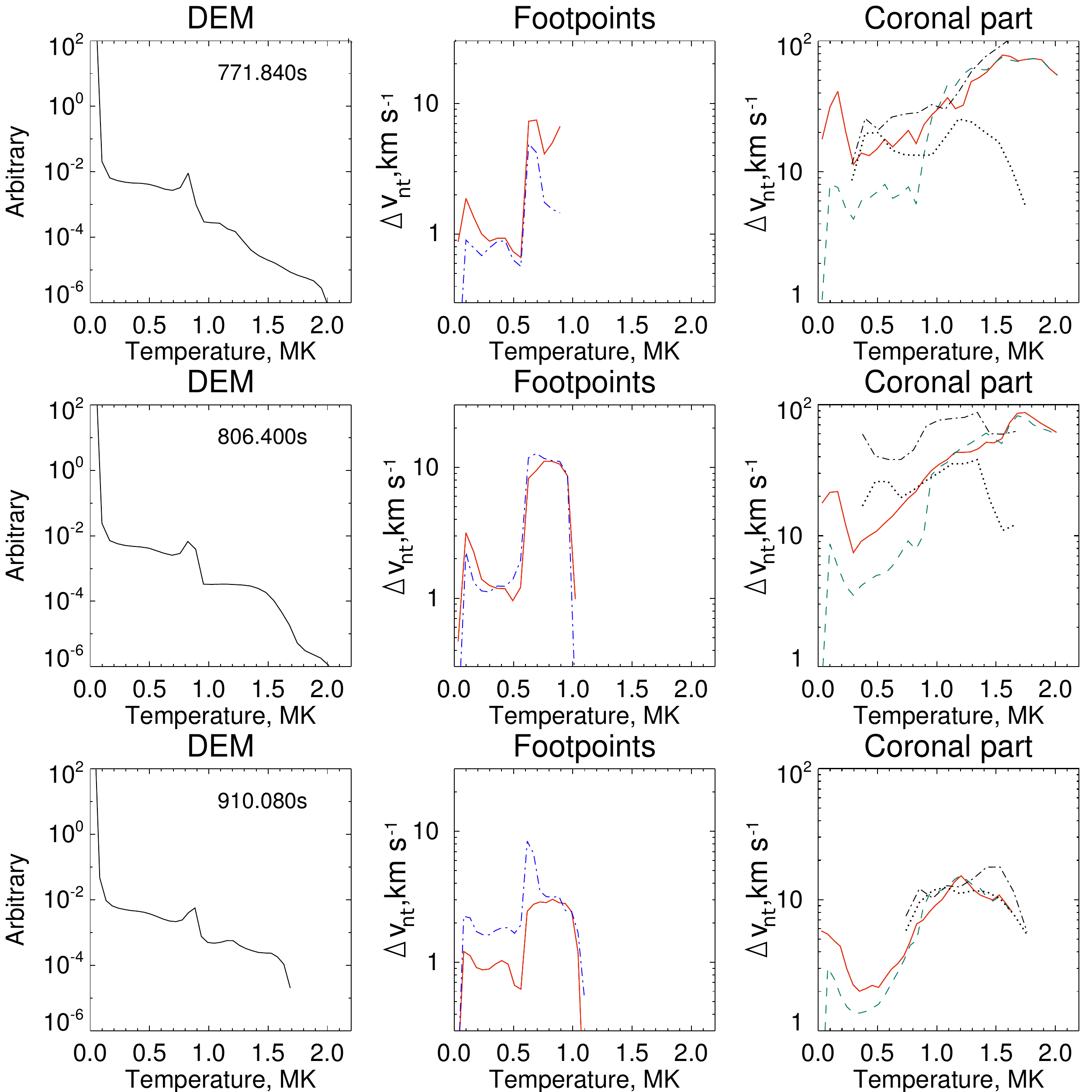}}
\caption{{\it Model S1}. {\it Left column:} Differential emission measure versus temperature for the whole domain. {\it Middle column:} Average $n^2$-weighted velocity dispersions $\Delta v_{nt}(T)$ (calculated as 
$\sqrt{\Delta v_{nt}^2(T)}$) 
for the foot-point location, for x-component (red solid lines) and z-component (blue dot-dashed lines) of the velocity.
{\it Right column:} 
Green dashed line is $\Delta v_{nt}(T)$ for the whole domain; red solid lines is the $\Delta v_{nt}(T)$ for the coronal part of the loop; black lines are $\Delta v_{nt}(T)$ for the loop-top region (dot-dashed line is for y-component and dotted line is for z-component of the velocity). Corresponding times are shown in the left panels. Locations of the sampling regions are shown in Figure~\ref{f-sketch}.}
\label{f-vels}
\end{figure*}

\begin{figure*}[ht!]    
\centerline{\includegraphics[width=0.8\textwidth,clip=]{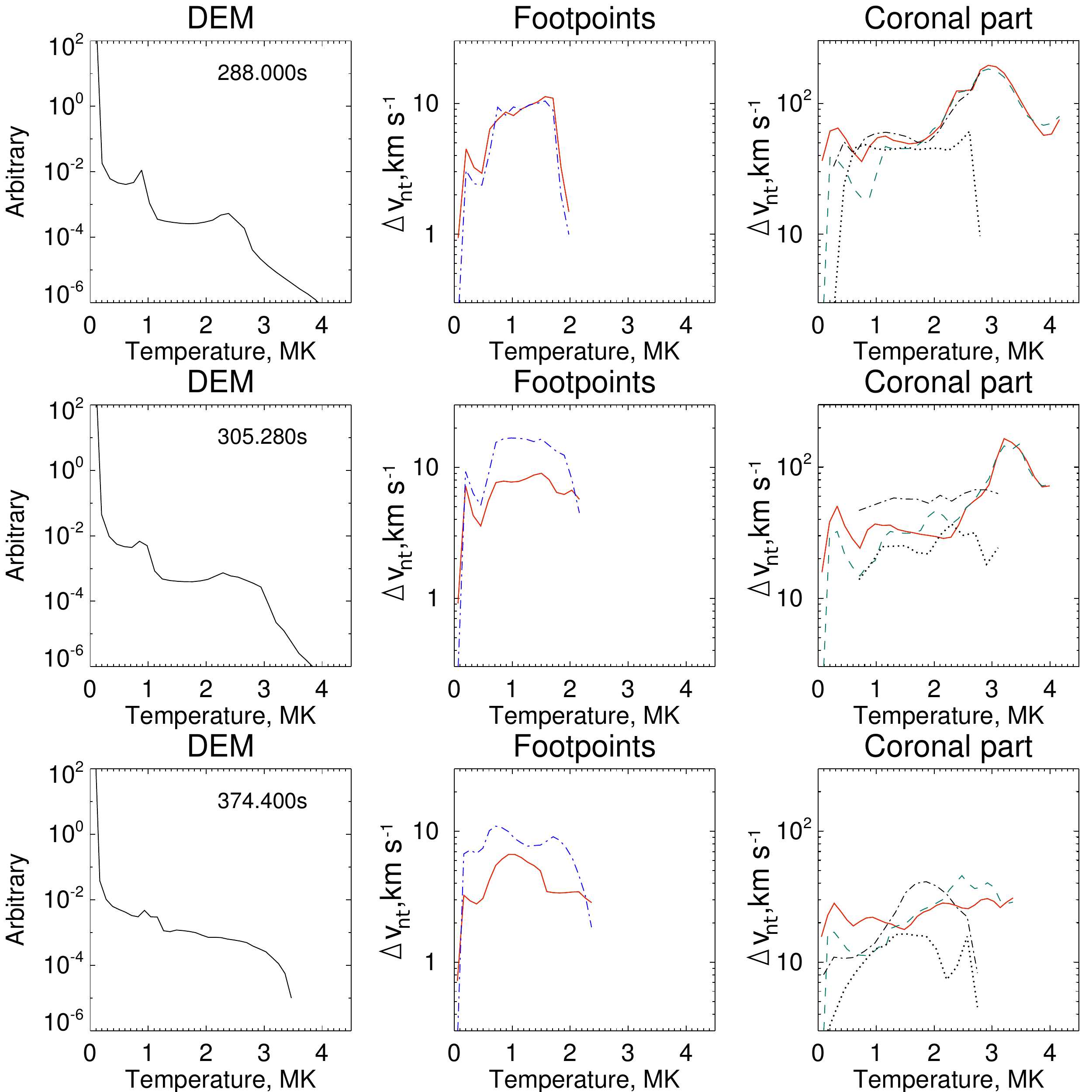}}
\caption{The same as in Figure~\ref{f-vels}, but for {\it Model S2}.}
\label{f-velv}
\end{figure*}

\begin{figure*}[ht!]    
\centerline{\includegraphics[width=0.8\textwidth,clip=]{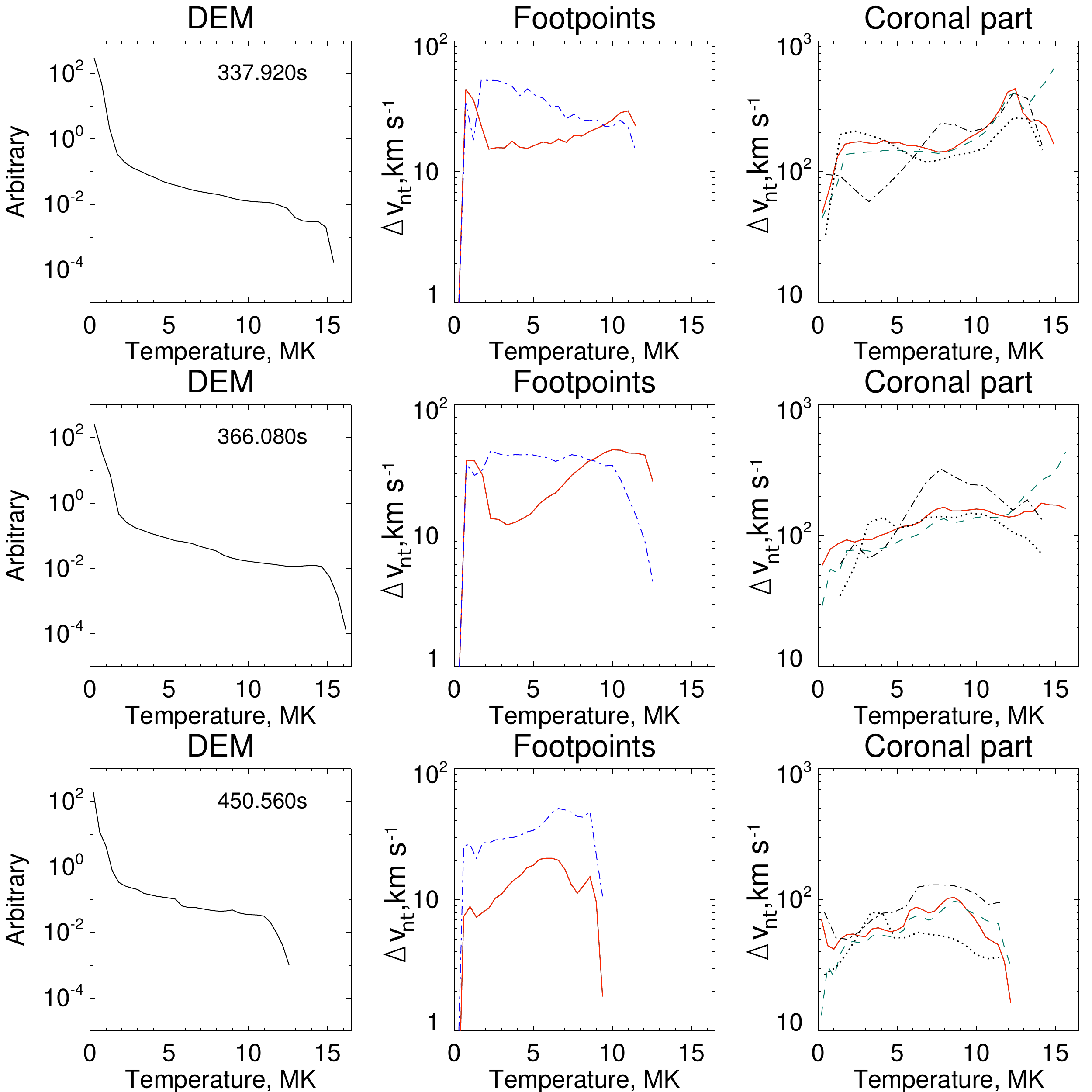}}
\caption{The same as in Figure~\ref{f-vels}, but for large-scale {\it Model L1}.}
\label{f-vely}
\end{figure*}

\begin{figure*}[ht!]    
\centerline{\includegraphics[width=0.8\textwidth,clip=]{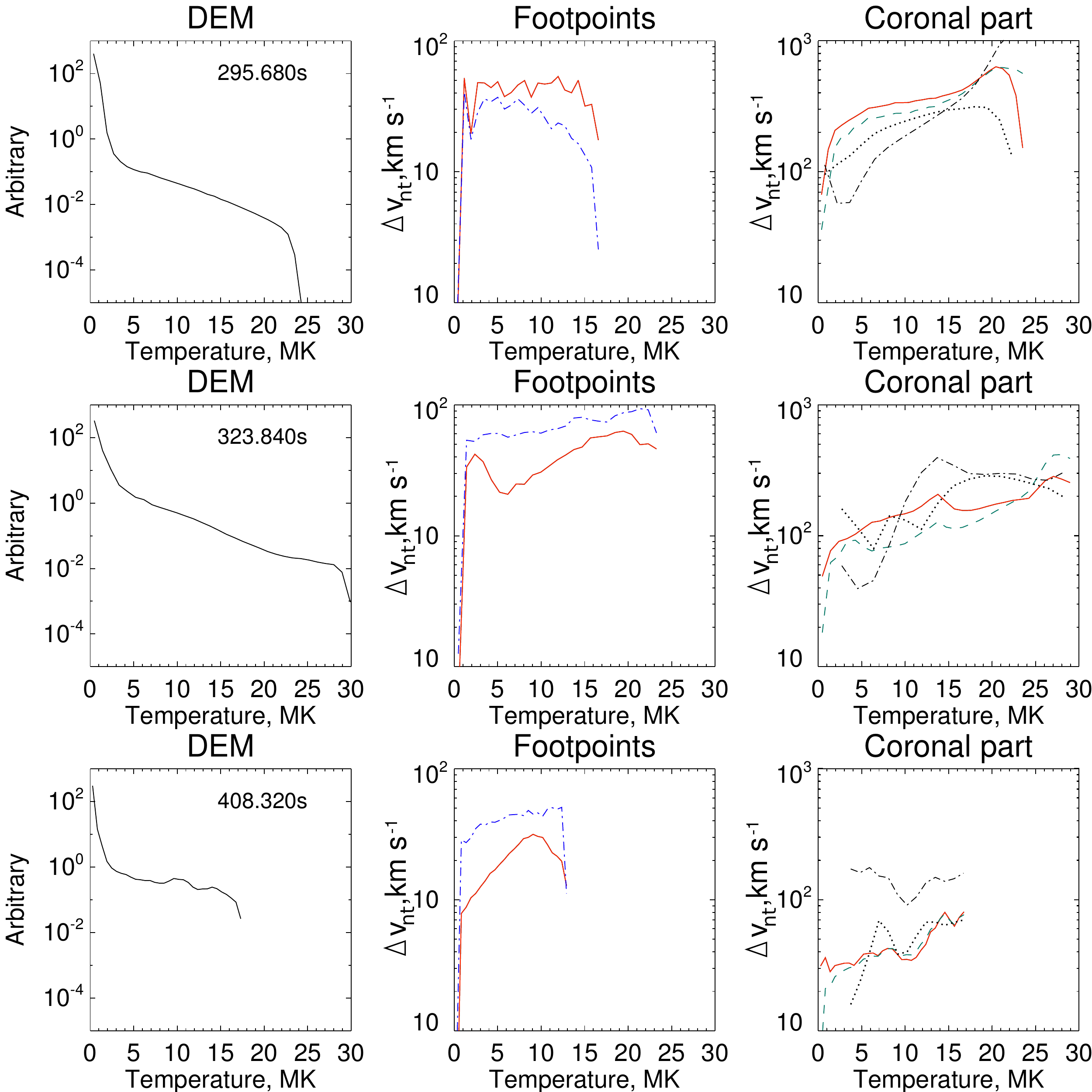}}
\caption{The same as in Figure~\ref{f-vels}, but for large-scale {\it Model L2}.}
\vspace{2 cm}
\label{f-velz}
\end{figure*}

\section{Unresolved plasma motions in different temperature ranges}\label{chars}

\begin{figure*}[ht!]    
\centerline{\includegraphics[width=0.5\textwidth,clip=]{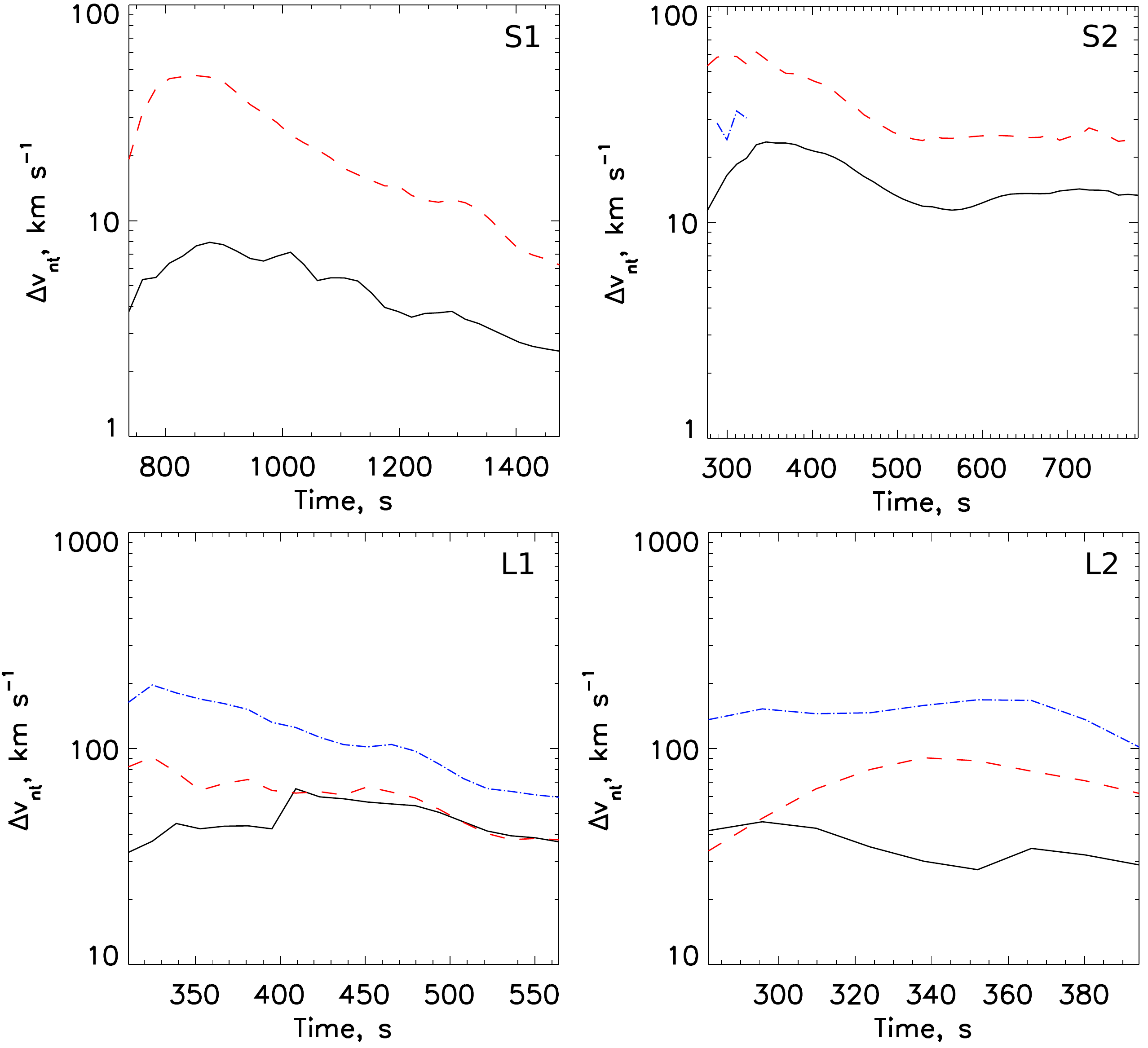}}
\caption{Variation of velocity dispersions calculated over the whole domain in three temperature bands in the four models after kink instabilities. Model names are shown in panels. Black solid line is for the $0.25$MK~$<\,T\,<$~$1$MK band, red dashed line is for $1$MK~$<\,T\,<$~$4$MK, and blue dot-dashed line is for $T\,>$~$4$MK. (There is practically no plasma with temperature $>4$~MK in models \it{S1} and \it{S2}.) Lower limits of each time intervale approximately correspond to the onset of kink instability and fast energy release.}
\label{f-velt}
\end{figure*}

\subsection{Macroscopic velocity distributions and their relation with non-thermal line broadening}\label{chars-eqs}

Here we derive expressions relating the resolved flow velocities and unresolved flow velocity distribution to widths and positions of spectral lines \cite[see e.g.][for exact derivations]{humi14}.

Assume the model contains only fully ionised hydrogen plasma, so that the electron and proton particle densities and total numbers are equal, $n_p=n_e=n$ and $N_p=N_e=N$ respectively. Further, assume that within an unresolved volume has a distribution of plasma in respect of the LOS velocity $v_{||}$ and temperature $T$, $f(v_{||},T) = \frac{d^2 N^2}{d v_{||} dT}$. Then, the profile of a spectral line emitted by this volume of plasma will be

\begin{equation}\label{eq-intens}
I(\lambda)=\int \limits_T \int \limits_{v_{||}} a(T) \frac{d^2 (N^2)}{d v_{||} dT} s\left(\lambda - \lambda_{line}-\frac {v_{||}}c \lambda_{line}\right) d v_{||} dT,
\end{equation}
where $a(T)$ is a temperature contribution function for the spectral line, $s(\lambda-\lambda_{line})$ is the line 
profile in the absence of macroscopic plasma motion.

The function $s$ is normally a Gaussian-like profile, which can be approximated by a Gaussian:
\begin{equation}\label{eq-thprof}
s(\lambda-\lambda_{line})= s_0 \exp\left(-\frac{(\lambda-\lambda_{line})^2}{\Delta\lambda_D^2}\right),
\end{equation}
where the Doppler width is $\Delta\lambda_D = \sqrt{\frac {2k_BT}{m_i c^2}}\lambda_{line}$. (In fact, the exact shape of the $s$ function is not important, as far as it is a Gaussian-like distribution, {\it i.e.} a distribution, becoming $0$ at $\pm \infty$, with finite dispersion.

Assuming that the function $\frac{d^2 N^2}{d v_{||} dT}$ is also Gaussian in respect of the LOS velocity, i.e.
\begin{equation}\label{eq-ntprof} \langle v \rangle
\frac{d^2 N^2}{d v_{||} dT} = \exp\left(-\frac{\left(v_{||}-\langle v \rangle (T)\right)^2}{\Delta v_{nt}^2(T)}\right) f_t(T),
\end{equation}
the Eq.~(\ref{eq-intens}) can be written (using the convolution of two Gaussian functions) as 
\begin{equation}
I(\lambda)=\int \limits_T \tau(T) \exp\left(-\frac{(\lambda-\lambda_{line}-\Delta \lambda_{LOS})^2}{\Delta \lambda_D^2 + \Delta \lambda_{nt}^2}\right) dT.
\end{equation}\label{eq-intens1}
Here $\Delta \lambda_{nt}=\frac{\Delta v_{nt}}c \lambda_{line}$ represents so-called non-thermal contribution to line broadening, while $\Delta \lambda_{LOS} = \frac {\langle v \rangle}c \Delta \lambda_{line}$, and $\tau(T)$ is the product of $a(T)$,$f_t(T)$ and constants.

Hence, if the plasma's distribution $\frac{d^2 N^2}{d v_{||} dT}$ for each small temperature interval in the considered volume has some Gaussian-like distribution in respect of $v_{||}$, then the average velocity, determining the line shift can be derived as follows:
\begin{equation}\label{eq-vavg}
\langle v_{||} \rangle(T) = \frac{\int \limits_V \frac{d^2 N^2}{d v_{||} dT}(T,v_{||})v_{||} dV}{\int \limits_V \frac{d^2 N^2}{d v_{||} dT}(T,v_{||}) dV}.
\end{equation}
(For this average LOS velocity we will use the notation $\langle v \rangle$ hereafter.)
The half-width of this distribution (or FWHM), which determines the non-thermal broadening of spectral line can be approximated as follows (this expression is exact for a Gaussian profile)
\begin{equation}\label{eq-vdis}
\Delta v_{nt}^2(T) = \frac {\int \limits_V \frac{d^2 N^2}{d v_{||} dT}(T,v_{||})(v_{||}-\langle v \rangle)^2 dV}{\int \limits_V \frac{d^2 N^2}{d v_{||} dT}(T,v_{||}) dV}.
\end{equation}
(For this LOS velocity dispersion (or variance) we will use the notation $\Delta v_{nt}$ hereafter.)

In our simulation we use discrete forms of Equations~\ref{eq-vavg} and \ref{eq-vdis}:

\begin{eqnarray}
\langle v \rangle_t &=& \frac {\sum \limits_\Omega [w_{ti} \; v_{||\,i} \; \delta\Omega]}{\sum \limits_\Omega [w_{ti} \; \delta\Omega]} \nonumber \\
\Delta v_{nt\;t}^2 &=& \frac {\sum \limits_\Omega [w_{ti} \; (v_{||\,i}-\langle v \rangle_t)^2 \; \delta\Omega]}{\sum \limits_\Omega [w_{ti}  \; \delta\Omega]} \nonumber ,
\end{eqnarray}
where 
\[
w_{ti} = \left\{
     \begin{array}{lr}
       \rho^2, & \mathrm{if}\; T_t < T_i < T_{t+1} \\
       0, & \mathrm{otherwise}
     \end{array}
   \right. \nonumber
\]

In these equations, the index $t$ corresponds to a temperature interval, index $i$ corresponds to a grid points, and $\Delta \Omega$ is an elementary volume between adjacent grid points (constant, as the grid is uniform in each direction). Here $w_{ti}$ is, effectively, the density of plasma squared, with temperature $T_t$ and velocity $v_{||\,i}$, and, hence, corresponds to $\frac{d^2 N^2}{d v_{||} dT}(v_{||}, T)$ in Equations~(\ref{eq-vavg}-\ref{eq-vdis}). The integration goes over all the grid points in the considered volume $\Omega$.

In the above equations, the integral
\[
M_t=\sum \limits_\Omega [w_{ti} \; \delta\Omega]
\]
is, in effect, the differential emission measure for the volume $\Omega$. (In observations it is calculated as $d[n^2(T) V(T)]/dT$ or $n^2 (dT/dl)^{-1}$ for an individual pixel.)

In section~\ref{chars-data} the velocities $\langle v \rangle$ and $\Delta v_{nt}$ are calculated as functions of temperature using the above equations and compared with observational data.

\subsection{Characteristics of the velocity fields obtained from numerical models}
\label{chars-data}

The velocity dispersion functions $\Delta v_{nt}(T)$ have been calculated for each model for four regions (see Figure~\ref{f-sketch}): large coronal region (practically, the whole domain volume above 2~Mm in {\it models S1} and {\it S2}, and above 8~Mm in {\it models L1} and {\it L2}), loop-top region (a cube with the size of 3~Mm in {\it S1} and {\it S2}, and 12~Mm in {\it L1} and {\it L2} containing the loop top), the footpoints (below 4~Mm in all four models), as well as for the whole domain. The values of $\Delta v_{nt}(T)$ are calculated separately using x-, y- and z-components of velocity. In the footpoint region, the z- and x- (or y-) componets of velocity correspond to directions along and across the magnetic field, respectively. At the same time, it is practically impossible to determine the dominant magnetic field direction in the large coronal sampling regions, because of the twist and fast fluxtube motions. However, we do this for the loop-top regions. Additionally, we calculate the Differential Emission Measure (DEM) functions for each model, in order to show the distribution of the hot radiating plasma. The results are shown in Figures~\ref{f-vels}-\ref{f-velz}.
The times used here are arbitrary, measured from re-starts of simulations. In each model, fast energy release begins from the onset of kink instability, which happens approximately approximately at 740~s in model {\it S1},  at 270~s in model {\it S2},  at 310~s in {\it L1}, and at 280~s in {\it L2}.

\begin{figure*}[ht!]    
\centerline{\includegraphics[width=0.8\textwidth,clip=]{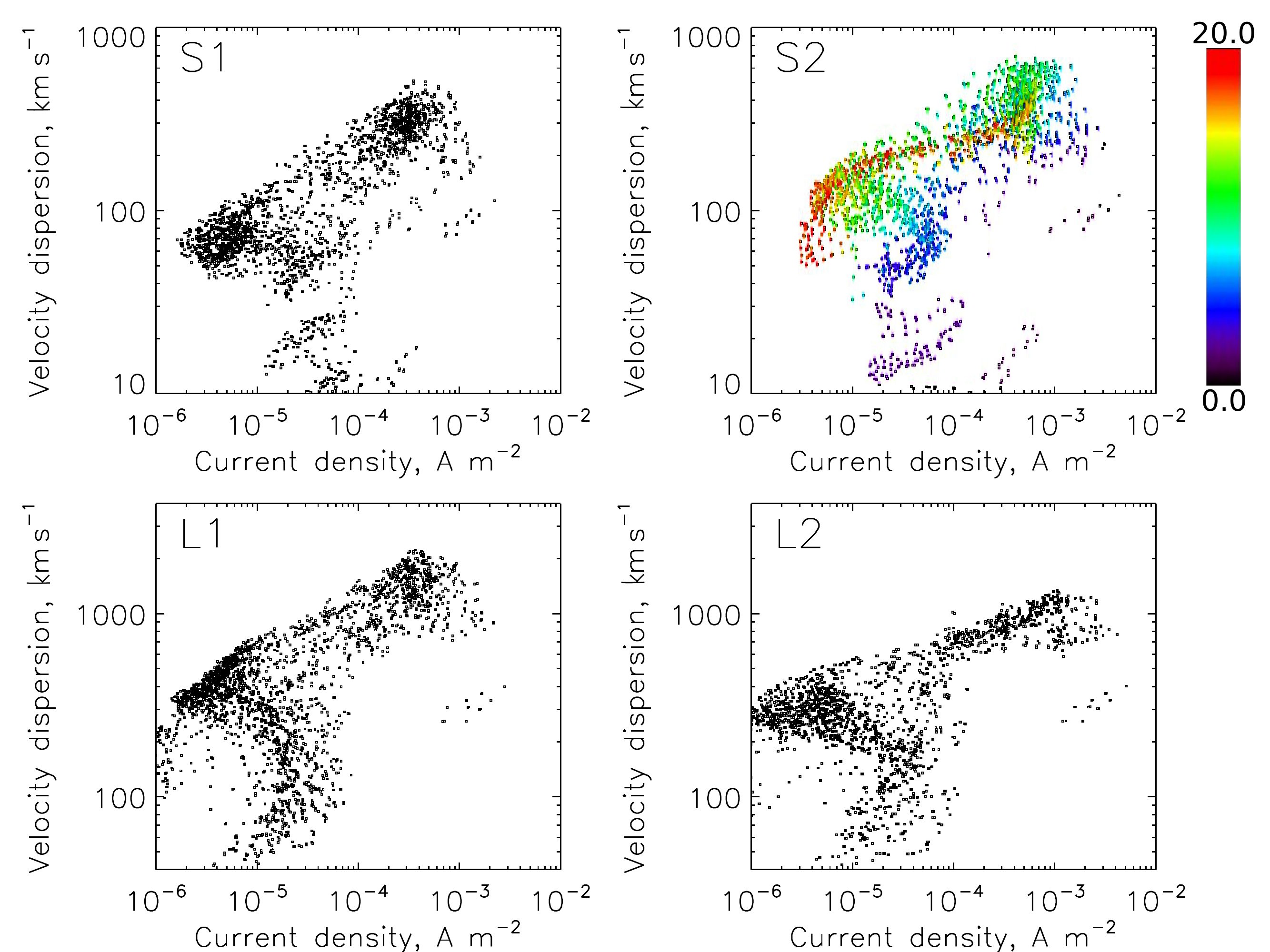}}
\caption{Average $n^2$-weighted velocity dispersions $\Delta v_{nt}$ versus current densities for the four models. Each dot represents a cube with the size of 2.5~$L_0$ (2.5~Mm in smaller models and 10~Mm in larger models). In order to demonstrate correlation with the height, colour flags indicating the height corresponding to each dot are shown in the panel S2.}
\label{f-vvj}
\end{figure*}

\begin{figure*}[ht!]    
\centerline{\includegraphics[width=0.7\textwidth,clip=]{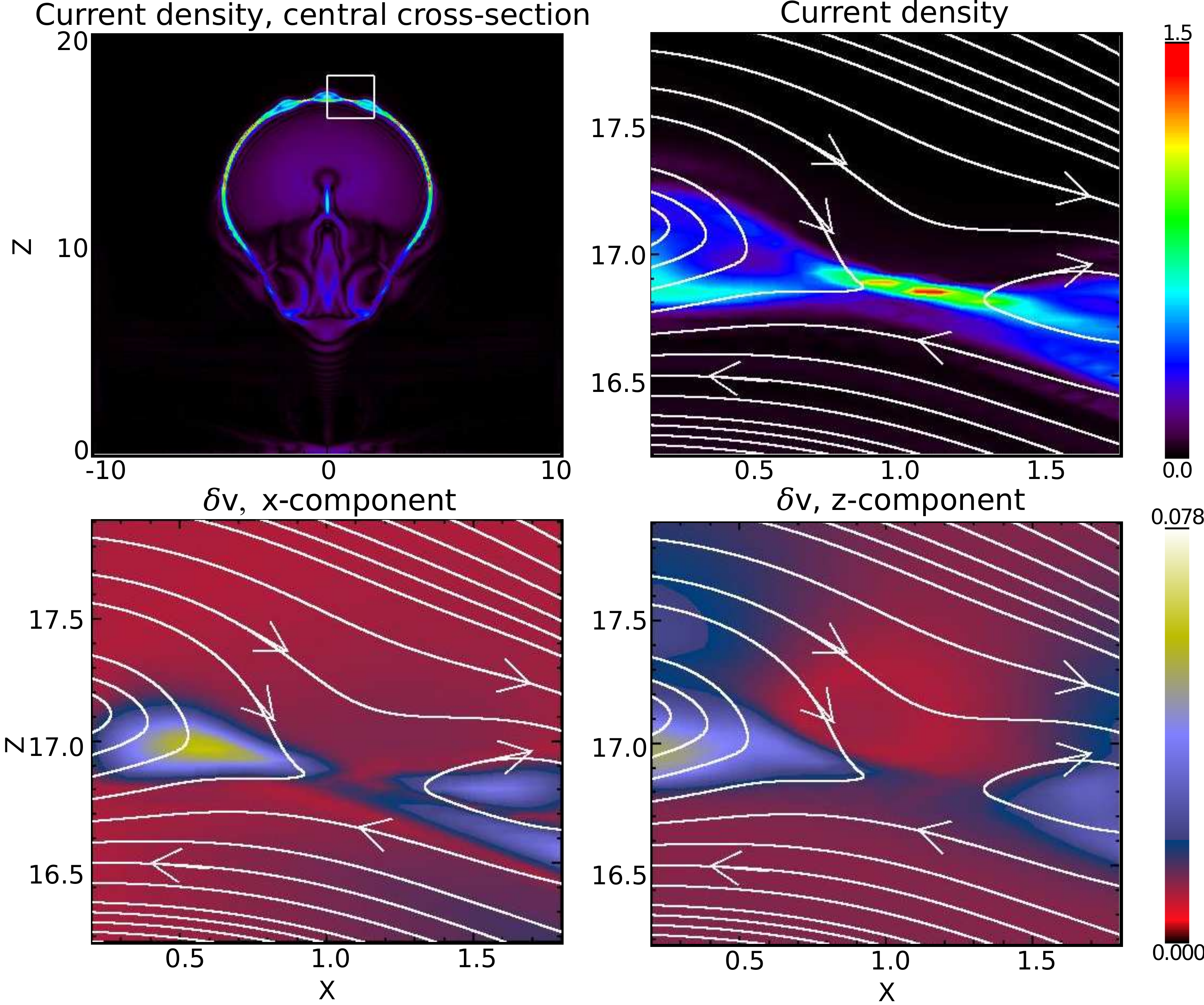}}
\caption{Distributions of current density $|j|$ and velocity deviations $\delta \vec{v}(x,z)$  around one of the energy release regions. Top left plot shows the global current density distribution at $x=0$ plane and the location of the sampled region. Current density distribution and magnetic field lines in the selected region are shown in the top right panel. Distributions of the $\rho^2 T$-weighted velocity deviations (see text for details) are shown in the bottom left (x-component) and bottom right (z-component) panels. (Here the units of the current density colourscale are $3.6\times 10^{-3}$~A~m$^{-2}$ and the units of the $\rho^2T$-weighted velocity colourscale are $2.8\times 10^3$~km~s$^{-1}$. Distances are in $10^6$~m.}
\label{f-vvloc}
\end{figure*}

\begin{figure*}[ht!]    
\centerline{\includegraphics[width=0.7\textwidth,clip=]{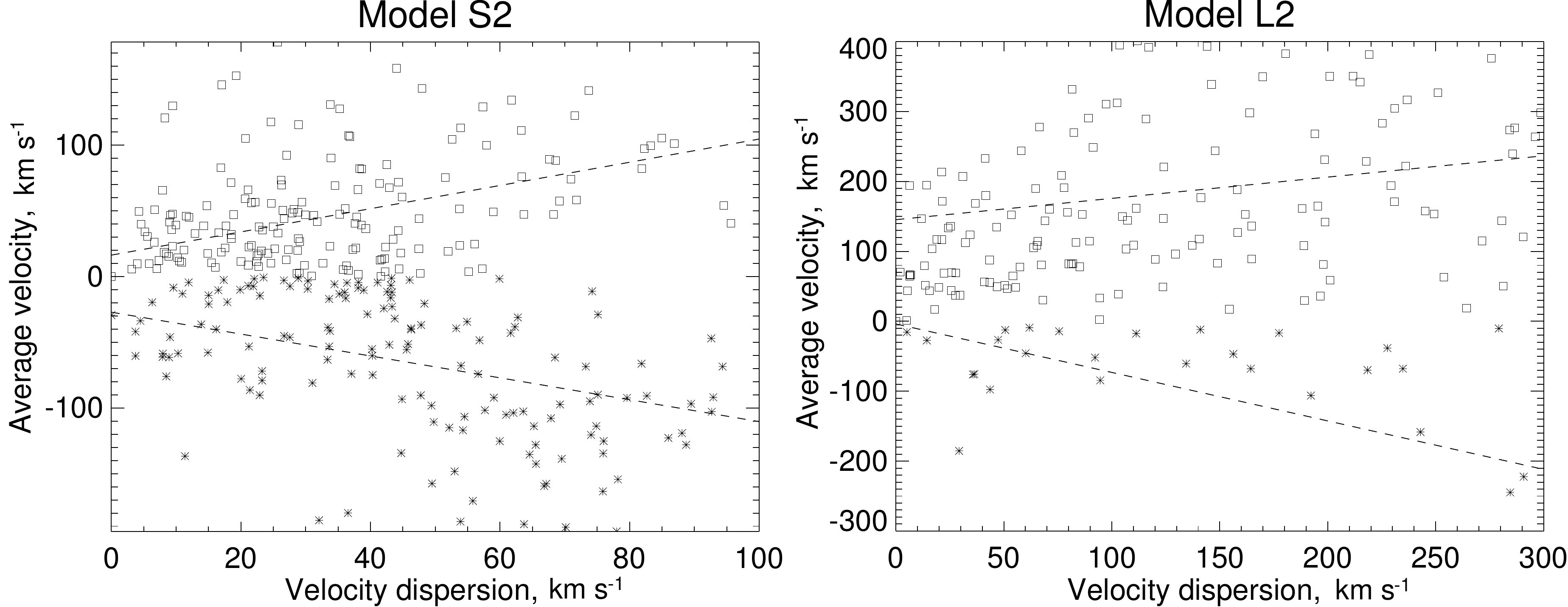}}
\caption{Average velocity ($Z$-component) versus velocity dispersion for {\it model S2} (left) and {\it L2} (right). Each point corresponds to a sampling cube with side 2.25~Mm in the small model (left) and 9~Mm in the large model (right). Here, positive velocities correspond to plasma moving up, and vice versa (contrary to the convention used in most observational papers).}
\label{f-vvv}
\end{figure*}

The DEM functions for small models (Figs.~\ref{f-vels}-\ref{f-velv}) can be divided into three parts. At low temperatures (~0.1~MK) there is a peak corresponding to the ``cold'' chromospheric plasma. It is followed by a nearly flat part and a peak at around 0.8-1.0~MK, corresponding to the coronal plasma. These two parts do not show substantial changes during loop evolution. The third part, at $T > 1$~MK, corresponds to the loop plasma heated during the reconnection. It is nearly flat, with slight positive slope during the reconnection phase and with slightly negative slope during the cooling phase. This part is limited by the peak temperature, which is about 2.5~MK for the {\it model S1} and about 3.5-4~MK for {\it S2}. DEM functions corresponding to the larger models (Figs.~\ref{f-vely}-\ref{f-velz}) can be divided into two parts: bulk of the plasma forms a peak with a strong negative slope at a few MK, and then show exponential profiles up to the maximum temperature. The maximum temperature reached during the reconnection phase in {\it model L1} is about 15-16~MK, while in the {\it model L2} it is about 30~MK.

Figures~\ref{f-vels}-\ref{f-velz} show LOS velocity dispersion as a functions of temperature, defined by Equation~\ref{eq-vdis}. The values of $\Delta v_{nt}$ change substantially between different phases: they are order of 10$^2$~km~s$^{-1}$ during the reconnection, and drop to $\sim$~10~km~s$^{-1}$ during the cooling phase. It can be seen that these values normally increase with the temperature in all regions in all considered models. Thus, the coronal parts of loops in small models demonstrate an increase in $\Delta v_{nt}$ from 10-30~km~s$^{-1}$ to about 100-200~km~s$^{-1}$ when the temperature increases from about 0.5-1~MK to 2-4~MK. In larger models, this value increases from 50-100~km~s$^{-1}$ at $T \approx$~2-3~MK to 200-400~km~s$^{-1}$ around 20~MK. Smaller coronal sampling regions show a similar picture. The situation near the footpoints is more complicated. Overall, it is possible to say that in most cases the velocity dispersion is noticeably higher at higher temperatures. Thus, typically, in {\it model S1} this velocity increases from few km~s$^{-1}$ below 1~MK to about 3-10~km~s$^{-1}$ at $\geq$1~MK. Similarly, in {\it model S2}, it also increases from few km~s$^{-1}$ just below 1~MK to about 10~km~s$^{-1}$ at $\sim$2~MK. In large models, this velocity is even higher; it increases from about 10~km~s$^{-1}$ at 2-3~MK to about 50-100~km~s$^{-1}$ at 10-15~MK. However, near foot-points the $\Delta v_{nt}(T)$ distribution sometimes is flat or even decreasing. 

Another interesting feature revealed by the non-thermal velocities near foot-points is the noticeable difference in dispersions in different directions. Thus, in the footpoints regions of smaller loops the $\Delta v_{nt}(T)$ distributions in x- and z-directions are similar at the beginning of reconnection. However, during the fastest stage of reconnection and during the cooling phase the dispersion of $v_z$ is 3-4 higher higher than the dispersion of 
$v_x$. Footpoints of larger loops show similar picture: both during the fast reconnection and during cooling, the velocity dispersions in vertical direction are 2-3 times higher than those in horizontal directions. In other words, the velocity dispersion along the magnetic field is higher than that across the field.

A similar picture can be seen in the small sampling region near loop tops: dispersions of $v_y$ (direction along the loop) are higher than dispersions of $v_z$ (across the loop). In smaller loops this is the case at three different stages, however, in larger loops this contrast is small during the onset of reconnection, becoming very substantial (up to 3 times) towards the end of reconnection. This also might be explained as the effect of the magnetic field, particularly, towards the end of reconnection, when the azimuthal field is substantially reduced. There is no noticeable difference in the velocity dispersions in different directions for the whole coronal part or for the whole domain. This is quite natural, taking into account that there is no substantial mean magnetic field in these models.

It should be noted that the plots in the middle column in Figures~\ref{f-vels}-\ref{f-velz} corresponding to the foot-point regions sample substantially large volumes (see Figure~\ref{f-sketch}). Their size is proportional to the domain (and loop) size and, therefore, in models {\it L1} and {\it L2} they sample larger volumes, including not only the chromosphere, but also the transition region and lower corona. This, most likely, explains why $\Delta v_{nt}(T)$ functions for foot-point and coronal regions are more similar in models {\it L1} and {\it L2}, compared to models {\it S1} and {\it S2}.

It should be noted that the non-thermal velocity dispersions near the footpoints, perhaps, are not representative of the whole loop: they correspond to a cooler and more structured material of the chromosphere and transition region, where plasma $\beta$ is higher and can be close to 1. However, they are still of interest, for instance, the variation of velocity dispersion with direction can indicate suppression or enhancement of the turbulence in relatively strong magnetic field. At the same time, the coronal part represents the whole loop: its $\Delta v_{nt}(T)$ distributions are very similar to those of the whole domain.  

Figure~\ref{f-velt} shows the time variation of the velocity dispersion functions in three different temperature ranges, 0.25~MK~--~1~MK, 1~MK~--~4~MK, and $>$4~MK. There is practically no signal in the highest temperature band in small models {\it S1} and {\it S2}. All four models show that $\Delta v_{nt}$ increases with temperature most of the time. Although, in model {\it L1} values of $\Delta v_{nt}$ in 0.25~MK~--~1~MK and 1~MK~--~4~MK bands are nearly equal towards the end of reconnection, while in model {\it L2} velocity dispersions in these bands are nearly equal just after the onset of reconnection.

In order to understand the relation between the primary energy release and velocity dispersions we calculated $\Delta v_{nt}$ as functions of the current density $j$ (Figure~\ref{f-vvj}). Similar to the temperature distribution of 
$\Delta v_{nt}$, these function were calculated using the original spatial resolution of our MHD simulations (about 0.07~$L_0$). Obtained $\Delta v_{nt}(j)$ functions did not show any noticeable spatial correlation between the velocity dispersions and current densities. However, when averaged over larger volumes (cubes with about 2.5~$L_0$ size), the velocity dispersions clearly show spatial correlation with current densities (Figure~\ref{f-vvj}). Moreover, there is also a correlation with height. Thus, there are two types of hot regions in the considered models. Coronal plasma (see red, yellow and green dots in Figure~\ref{f-vvj}, panel S2) shows weaker increase of the velocity dispersion with the current density ($\Delta v_{nt} \sim j^{0.35}$), while the hot regions in footpoint show much steeper increase of $\Delta v_{nt}$ with currents ($\Delta v_{nt} \sim j$). 
The effect of spatial resolution here can be explained by the structure of the primary energy release sites. Figure~\ref{f-vvloc} shows the (x,z) distributions of current density and the velocity deviations, as well as magnetic field lines around a region with high current density at the 'current shield' formed just above the top of a twisted loop. Here, velocity deviations are calculated as 
\[
\delta v_r = \sqrt{(v_r (x,z)-v_{r\;mean})^2\frac{\rho^2(x,z) T(x,z)}{\rho^2_{mean} T_{mean}}},
\]
where index $r$ means $x$, $y$ or $z$, and $\rho_{mean}$ and $T_{mean}$ are the density and temperature averaged over the sampled region, respectively.
This region has all basic features of reconnection current sheet, including the inflow and outflow regions \cite[see also][]{bare15}. Importantly, plasma with high turbulent velocities is located in the exhaust regions, away from the current layer, and, hence, there is no immediate spatial correlation between the turbulent velocities and current density. However, the values of $\Delta v_{nt}$ averaged over volumes larger than these elementary current sheets will show spatial correlation with current densities.

On the other hand, relatively weak correlation and a substantial spread of points in the $\Delta v_{nt}$ versus $|j|$ diagrams means that plasma turbulence and strong current are not always connected. Firsly, this is because there are several factors, which are not directly related to the energy release, can result in velocity dispersion (see Sect.~\ref{chars-obs}). Secondly, the magnetic energy can be converted into kinetic energy of plasma turbulence well away from the reconnection regions. This is possible, for instance, if the plasma is heated and turbulised by MHD shocks generated by the magnetic reconnection \cite{baho15,bare15}.

We also compare the velocity dispersion (related to line width) with the bulk flow velocities (related to line shift); the measurements for our two models, {\it S2} and {\it L2} are shown in Figure~\ref{f-vvv}. They are calculated using equations~\ref{eq-vavg}-\ref{eq-vdis} and integrated over temperature. Only plasma with temperature higher than 1~MK is taken into account here. Each point corresponds to a cube with the side of about 30 grid points (2.25~Mm in {\it model S2} and 9~Mm in {\it L2}). It is found that the average velocities are rather low when the non-thermal velocity dispersion is low, $\langle v \rangle\approx$50~km~s$^{-1}$ when $\Delta v_{nt}=$20~km~s$^{-1}$. Then, it increases with $\Delta v_{nt}$, reaching around $\langle v \rangle=$80-100~km~s$^{-1}$ for $\Delta v_{nt} \approx$100~km~s$^{-1}$, and about $\langle v \rangle=$700~km~s$^{-1}$ for $\Delta v_{nt} \approx$1000~km~s$^{-1}$. Both negative and positive average velocities are present, and the dependences between $\Delta v_{nt}$ and $\langle v \rangle$ can be linearly approximated both for upflows and downflows. Although these dependences are clearly visible on the graphs, the spread of points is very big (several km~s$^{-1}$).

\subsection{Comparison with observations and discussion}\label{chars-obs}

Our models can reproduce the two key observational features described in Section~1. Firstly, similarly to observational data, the non-thermal velocity dispersion in the model increases with temperature, at least in its coronal part (which is the most of the volume for most loops). The exact values depend substantially on the model (or loop) size. At temperatures of about 1~MK, these velocity dispersions are about 10-50~km~s$^{-1}$, increasing to about 100-200~km~s$^{-1}$ at temperatures of 5-10~MK and reaching up to 500~km~s$^{-1}$ in the range of 20-30~MK. Observational data (see Sect.~1) similarly shows an increase from 10-20~km~s$^{-1}$ at $\sim$1~MK to few hundreds of km~s$^{-1}$ at about 10~MK. Also, similar to observational data, our models show a noticeable peak in $\Delta v_{nt}(T)$ functions at low temperatures, around 0.3-0.5~MK.

Secondly, the loops considered in our models also demonstrate correlation of the bulk velocity with non-thermal velocity dispersion: the higher the LOS flow velocity, the higher is the $\Delta v_{nt}$ value. Both, downflows and upflows are present (nearly equally). However, the average LOS velocities derived from our models demonstrate a substantial spread of about 100~km~s$^{-1}$, while in the observations the $\langle v \rangle$ distribution is more compact \citep{dose08}. This spread is larger in the plasma with positive velocities (upflow). Finally, the range of average LOS velocities obtained in our simulations is considerably larger (up to 800~km~s$^{-1}$) than in observations by Doschek, mentioned above. However, these values are still acceptable, as such velocities are observed sometimes in larger loop \citep{cire13}.

The correlation between temperature and velocities is not surprising: indeed, one would expect that plasma with more heating would contain more kinetic energy in form of large-scale turbulence and regular flows. Thus, in our simulation highest velocities appear in and around the energy release regions (close to the loop top, and in loop legs closer to current concentrations near footpoints). This also explains the correlation between the velocity dispersion and the average flow velocity. Indeed, both values $\langle v \rangle$ and $\Delta v_{nt}$ are high in the hot plasma regions, relatively low in colder regions of the flaring loop, and quite low outside the loop. Generally, it means that 
a single energy release mechanism is responsible for small-scale turbulent velocities and large-scale plasma motion, so that the energies of these two types of motion are proportional to each other.
The reason why the spread in velocities obtained in our simulations (Figure~\ref{f-vvv}) is substantially greater than in observations is less clear. This may be because the large box we use in our measurements includes very different regions. On the other hand, it may be due to the fact, that in observations the values of $\langle v \rangle$ and $\Delta v_{nt}$ are calculated over a larger temperature range than in our simulations.

Concerning the velocity dispersion, it is important to note that there are several different factors causing them, and some of these factors may be unrelated to the energy release. The key factor, which is in the focus of this paper, is the turbulence in the very hot flaring plasma. As shown in the Section~\ref{chars-data}, this turbulence occurs in the very hot plasma ejected from elementary reconnecting current sheets. Since the reconnection is the most powerful heating mechanism in our models, this mechanism would be dominant at high temperatures (1 MK and higher). Secondly, there are fast chaotic motions due to the loop kinking after the instability. Since these velocities should be highest close to the loop top, loop kinking would affect the temperature range $>$0.5~MK ({\it i.e.} the temperature of the quiet coronal plasma and the temperature of flaring plasma. There is also an important implication from this mechanism: noticeable non-thermal velocity dispersions may appear even before the heating, as it is found in observations \citep{hare13}. Thirdly, the thermal flux from the corona can heat dense transition regions and the chromosphere, which may add to the velocity dispersion at lower temperatures 10$^4$--10$^6$~K via non-uniform evaporation upflows. 

Substantial difference in the velocity dispersions in different directions can be explained assuming that the mean magnetic field is close to the loop direction. (Here, by the loop direction we mean the ``skeleton line'' of a loop going from one footpoint to another footpoint via the loop top.) This would mean that the velocity dispersion along the mean magnetic field is higher than across it. There are two factors, that could lead to this. Firstly, the turbulence in the hot flaring plasma can be anisotropic, being stronger along the mean field. Secondly, since the hot plasma is moving predominantly along magnetic field, any cross-field inhomogeneity of the flow velocity at small spatial scales will result in the line-of-sight velocity dispersion, if observed along the magnetic field direction. This inhomogeneity is quite possible because of the anisotropic thermal conduction: heat is quickly redistributed along magnetic field lines with almost no conduction in the perpendicular direction.

Finally, there are two features in the velocity field unrelated to the instability and the energy release: foot-point rotation and kink-oscillations of the loops before the instability. Thus, slow periodic variation in the total kinetic energy in these loops \cite[see e.g. Fig.4 in][]{gore14} are likely to be due to low-amplitude fundamental mode ({\it i.e.} $N=1$) kink oscillations. Based on the kinetic energy variation, their amplitude should be order of $0.001 v_0$ or smaller ({\it i.e.} about 1~km~s$^{-1}$ or smaller). Hence, these oscillations, in principle, could make a small contribution to the velocity dispersions, particularly at low temperatures. 
The maximum linear speed due to the loop rotation is about $0.004 v_0$, or 10~km~s$^{-1}$. This motion also can contribute to the velocity dispersion, especially at low temperature (since the velocity dispersion is $n^2$-weighed and the temperatures in the dense chromosphere are $10^4$--$10^5$~K. However, the velocity dispersions higher than $\sim 10$~km~s$^{-1}$ cannot be attributed to the driver.

\section{Summary}\label{summ}

In the present work we derive characteristics of the velocity field in flaring loop and compare them with observations, focusing on the plasma velocity dispersions, which correspond to non-thermal broadening of coronal EUV lines. It is found that our models yield velocity dispersions and average line-of-sight velocities, which are in qualitative and quantitative agreement with those derived from observational data.

Thus, velocity dispersions $\Delta v_{nt}$, average line-of-sight velocities $\langle v \rangle$ and maximum temperatures depend on the flare size and characteristic magnetic field in the loop. Thus, in small loops (with length of about 20~Mm) the maximum temperature varies between 2~MK (loop with footpoint magnetic field about 100~G) and 4~MK (footpoint field 200~G), average velocities and velocity dispersions are about 10-100~km~s$^{-1}$. In large loops (with length of about 80~Mm) the temperature can reach 16~MK (footpoint field 700~G) and 30~MK (footpoint field 1.5~kG), while the average velocities and velocity dispersions are 50-500~km~s$^{-1}$.

In the chromospheric and photospheric plasma close to footpoints the velocity dispersion is low, about 5-20~km~s$^{-1}$. The velocity dispersion along the field is found to be lower, by factor of 2-4, than that across magnetic field.

The velocity dispersion correlates with temperature. Thus, it is around 10~km~s$^{-1}$ near $T\approx$~0.5-1.0~MK, increases to about 100~km~s$^{-1}$ at $\sim$2~MK, and reaches about 200-500~km~s$^{-1}$ at $T\approx$~10-20~MK.

The velocity dispersions show spatial correlation with current densities, they vary approximately as $\Delta v_{nt} \sim j^{0.35}$ in the corona, and as $\Delta v_{nt} \sim j$ near footpoints.

Velocity dispersions appear to be higher along the loop direction, both near the footpoints and at the loop tops. Typically, the ratio is about 2--4.

Finally, the velocity dispersion correlates with average LOS velocity. Average velocities can be both positive and negative (upflow and downflow). At lower $\Delta v_{nt}$ values these two quantities are connected approximately as $\langle v \rangle \approx \Delta v_{nt}$, while at high values $\Delta v_{nt} \geq 100$~km~s$^{-1}$ this relation is $\langle v \rangle \approx 0.3 \Delta v_{nt}$ for upflow and $\langle v \rangle \approx 0.6 \Delta v_{nt}$ for downflow.

Therefore, based on our simulations, we can conclude that the correlation between the observed velocity dispersions (derived from spectral line widths) and the temperature, as well as the correlation between the average LOS velocities (derived from spectral line shifts) and the velocity dispersions, indicate that a single mechanism, direct plasma heating during magnetic reconnection, is responsible both for turbulisation of plasma and bulk plasma motion.

It is important to say that a good agreement with observations doesn't mean that all flares investigated in the observational studies (mentioned in Sect.~\ref{intro}) occurred in twisted coronal loops. However, the fact that solar flares occurring in this type of magnetic configurations produces data similar to observations implies that the 'twisted loop' configuration is a viable model for a solar flare, which can explain observations of $\langle v\rangle$ and $\Delta v_{nt}$ using simple reconnection scenario with multiple extended reconnection regions.

\begin{acknowledgements}
This work is funded by Science and Technology Facilities Council (UK) (STFC). MHD simulations have been performed using UKMHD facilities in the University of St Andrews and DIRAC system at Durham University, operated by the Institute for Computational Cosmology on behalf of the STFC DiRAC HPC Facility (www.dirac.ac.uk). DiRAC is part of the National E-Infrastructure.
\end{acknowledgements}

\end{document}